\documentclass[aps,pra,groupedaddress,notitlepage, showkeys]{revtex4-1}

\usepackage{amssymb}
\usepackage{amsmath}
\usepackage{amsfonts}
\usepackage{amsthm}
\usepackage{graphicx}
\usepackage{xcolor}
\usepackage{bm}
\usepackage{url}
\usepackage{hyperref}
\usepackage{cases}

\begin{document}

\title{Analog-based ensembles to characterize turbulent dynamics from observed data}

\author{Carlos Granero-Belinch\'on} 
\affiliation{Department of Mathematical and Electrical Engineering, IMT Atlantique, Lab-STICC, UMR CNRS 6285, 655 Av. du Technop\^ole, Plouzan\'e, 29280, Bretagne, France.}
\affiliation{Odyssey, Inria/IMT Atlantique, 263 Av. G\'en\'eral Leclerc, Rennes, 35042, Bretagne, France.}
\email{carlos.granero-belinchon@imt-atlantique.fr}

\date{\today}

\begin{abstract}
We present a methodology for the study of the dispersion of trajectories of stochastic processes in reconstructed phase spaces from observed data. The methodology allows to find ensembles of analog states, \textit{i.e.} states that are close in the phase space. Once these states are found, we focus on the characterization of their dispersion in function of 1) the time and 2) their initial separation. We study an experimental turbulent velocity measurement and two scale-invariant stochastic processes: a regularized fractional Brownian motion and a regularized multifractal random walk. Both stochastic processes are synthesized to have the same covariance structure as the experimental turbulent velocity, but only the regularized multifractal random walk mimics the intermittency of turbulent velocity. We illustrate that while the covariance structure of the processes governs the time dependence of the dispersion of the analog states, the intermittency phenomenon is responsible for the impact of the initial separation of the analogs on their dispersion. 
\end{abstract}


\maketitle

\section{Introduction}

Chaotic systems are characterized by complex and unpredictable behavior despite being governed by deterministic laws~\cite{Cencini2009, Eckmann1985, Lorenz1963}. This is due to an extreme sensitivity to initial conditions that can be quantified through the Lyapunov exponent. A positive Lyapunov exponent indicates an exponential rate of separation of infinitesimally close trajectories in phase space~\cite{Cencini2009,Lyapunov1992}:

\begin{equation}\label{}
|\delta(t+\tau)| \approx |\delta(t)| e^{\lambda \tau}
\end{equation}

\noindent where $|\delta(t)|$ is the initial separation of the trajectories at time $t$, $|\delta(t+\tau)|$ is the separation of the trajectories at time $t+\tau$, and $\lambda$ is the Lyapunov exponent, with $\lambda>0$ being the signature of chaos. For any finite $\lambda$ and for any $\tau > 0$, $|\delta(t+\tau)|$ goes to $0$ when $|\delta(t)|$ goes to zero, \textit{i.e.} two trajectories starting at the same exact location of the phase space evolve identically. 
These results are valid for smooth systems~\cite{Eckmann1985,Benettin1980,Benettin1980a} \textit{i.e.} those described by Lipschitz continuous functions~\cite{Bernard1998,Falkovich2001}. These systems offer the advantage of well-defined mathematical structures that allow for a rigorous treatment of their dynamical properties. 

Turbulence is a multiscale phenomenon, involving non-linear interactions across scales that lead to an energy cascade from large scales where energy is injected, down to small scales where it is dissipated. In the case of homogeneous and isotropic turbulence, this process defines three distinct domains of spatial scales. First, the integral domain, which includes scales larger than the integral scale $L$, where energy is injected into the system. Second, the inertial domain, encompassing scales smaller than $L$ but larger than the dissipative scale $\eta$, where energy is transferred from larger to smaller scales. Lastly, the dissipative domain, consisting of scales smaller than $\eta$, where the energy is dissipated~\cite{Frisch1995}. Each spatial scale has an associated time scale. Thus, we can define an integral, $T$, and dissipative, $\tau_K$, time scales that are associated to the dynamics of structures of size $L$ and $\eta$ respectively~\cite{Chevillard2005}. 

For scales in the dissipative domain, the turbulent velocity field is considered smooth. However, from Kolmogorov K41 theory~\cite{kolmogorovLocalStructureTurbulence1941a,kolmogorovLocalStructureTurbulence1991,Frisch1995} at scales in the inertial domain, the turbulent velocity field is not differentiable but only H\"older continuous with H\"older exponent $h=1/3$. Actually, more recent theories of turbulence state that the velocity field is multifractal~\cite{Kolmogorov1962,Obukhov1962,frischSingularityStructureFully1985,Frisch1995}. It possesses a range of scaling exponents ($h_{\text{min}},h_{\text{max}}$) with each $h$ in this range appearing in a different spatial set $\mathcal{S}_h$ of fractal dimension $D(h)$. This implies that the turbulent velocity field becomes smooth at different small scales depending on the spatiotemporal region~\cite{Paladin1987}.

The non-differentiability of turbulent velocity leads to an intrinsic probabilistic character that has been extensively illustrated through the numerical study of Lagrangian tracers~\cite{Biferale2005, Eyink2011, Thalabard2014, Thalabard2020}: two Lagrangian particles advected by a turbulent flow present, at sufficiently large times $t$, a separation that is independent of the initial separation of the particles~\cite{Richardson1921,Eyink2011}. This independence on the initial separation, that is a direct consequence of the H\"older continuity of the velocity field, leads to $|\delta(t+\tau)| \neq 0$ when $|\delta(t)|=0$ \textit{i.e.} particles initially at the same position separate when the time evolves. This was shown by Bernard et al.~\cite{Bernard1998} for Lagrangian tracers advected in Holder continuous fields in space and decorrelated in time. 

A lot of works have demonstrated the intrinsic probabilistic character of the Lagrangian turbulent flow~\cite{Biferale2005, Thalabard2014, Thalabard2020}. This phenomenon has been also called: spontaneous stochasticity~\cite{Falkovich2001,Eyink2011} or intrinsic stochasticity~\cite{Weinan2000}. These works are based on the study of the dispersion of Lagrangian particles that are close in space, and so they do not directly deal with trajectories in the phase space. There exist also some works trying to generalize this intrinsic probabilistic character to Eulerian turbulent flows~\cite{Bandak2024}, but they are based on simplified models of turbulence since an ensemble of realizations of the flow sharing the same, or infinitesimally close, initial condition is needed. All these studies are mainly based on numerical simulations, and experimental evidences are still unclear~\cite{Bourgoin2006}. 

We propose to reconstruct a phase space of turbulent velocity from 1-dimensional experimental measurements using Takens embedding~\cite{Takens1981} and to study the dispersion of close trajectories in this reconstructed phase space. We adapt analog-based methods\cite{lorenzAtmosphericPredictabilityRevealed1969a} previously used in meteorology \cite{doolNewLookWeather1989a,tothLongRangeWeatherForecasting1989} and dynamical systems~\cite{Sugihara1990} to generate ensembles of trajectories of turbulent velocity that are close in the reconstructed phase space. Once these ensembles are generated, we study the separation $|\delta(t+\tau)|$ between trajectories in function of 1) the time delay $\tau$ and 2) their initial separation $|\delta(t)|$. To isolate the contributions of the different properties of turbulence to the dynamics of the trajectories, we also study two scale-invariant stochastic processes mimicking diverse statistical properties of turbulence: a regularized fractional Brownian motion (r-fBm) and a regularized multifractal random walk (r-MRW)~\cite{Robert2008,Chevillard2010}. Both present a covariance structure perfectly matching the one of turbulent velocity, but only the r-MRW has intermittency, \textit{i.e.} is multifractal presenting extreme events at small scales. 

We obtain two main results:

\begin{itemize}
\item the separation of trajectories has the same time dependence for the experimental turbulent velocity and both stochastic processes. It presents three different regimes: for time delays $\tau$ smaller than the Kolmogorov dissipative scale $\tau_K$ the separation of trajectories behaves as $\tau^2$, for times larger than the integral scale $T$ the separation of trajectories remains constant (trajectories cover the full phase space), for times larger than the Kolmogorov dissipative scale and smaller than the integral scale $T$, the separation between trajectories behaves as $\tau^{2/3}$. 

\item Intermittency introduces a dependence of $|\delta(t+\tau)|$ on $|\delta(t)|$. In the case of r-fBm, that is a monofractal process with $h=1/3$ (non-intermittent), for any $\tau > \tau_K$ the separation of trajectories is independent of their initial separation. However, for r-MRW and the experimental turbulence, that are multifractal with a full set of scaling exponents $h$ (intermittent), the separation of trajectories behaves as a power law of their initial separation.
\end{itemize}

The article is organized as follow. Section~\ref{sec:method}, describes the analog-based methodology used to generate ensembles of states of a stochastic process $X$, and how these ensembles are employed to study the separation of trajectories in the reconstructed phase space. In section~\ref{sec:data}, we present the experimental turbulent velocity measurement and the two scale-invariant stochastic processes analyzed in this study. In section~\ref{sec:num}, we detail the numerical implementation of the analog-based ensemble generation, including the definition of the databases used to search for analogs and the selection of algorithm parameters. In section~\ref{sec:results}, we analyze, for the turbulent experimental velocity, the r-fBm and the r-MRW, the dispersion of trajectories in the phase space in function of their initial separation and the time. Finally, section~\ref{sec:conclusion} presents some conclusions and perspectives.

\section{Analog-based generation of ensembles for phase space sampling}
\label{sec:method}

Given a 1-dimensional stochastic process $X=\{x(t)\}$ homogeneously sampled at intervals $dt$, we consider Taken's embedding~\cite{Takens1981} to produce temporal sub-sequences $\vec{x}^{(p)}(t)$ of size $p$ at time $t$:

\begin{equation}
    \vec{x}^{(p)}(t) = 
    \begin{pmatrix}
        x(t) \\
        \vdots \\
        x(t-(p-1)dt)
    \end{pmatrix}
\end{equation}

\noindent where $(p-1) dt$ is the length of the observed past and $p$ defines the dimension of the reconstructed phase space~\cite{Takens1981, Eckmann1985, Abarbanel1992}.

Two such sub-sequences $\vec{x}^{(p)}(t)$ and $\vec{x}^{(p)}(t')$ at two different times $t$ and $t'$ are \emph{analogs} with respect to a given distance definition $D$, if $D(\vec{x}^{(p)}(t), \vec{x}^{(p)}(t')) < \epsilon$, where $\epsilon$ is a threshold. In this work, we consider the Euclidean distance, as suggested by~\cite{platzerUsingLocalDynamics2021c,lguensatAnalogDataAssimilation2017a}. Each sub-sequence $\vec{x}^{(p)}(t)$ has an associated \emph{successor} $\vec{x}^{(q)}(t+\tau)$, which is nothing but the $q$-dimensional sub-sequence of the stochastic process taken $\tau$ time later. This definition of successors implies a reconstructed $q$-dimensional phase space of the successors.

\begin{figure}[ht]
 \includegraphics[width=1\linewidth]{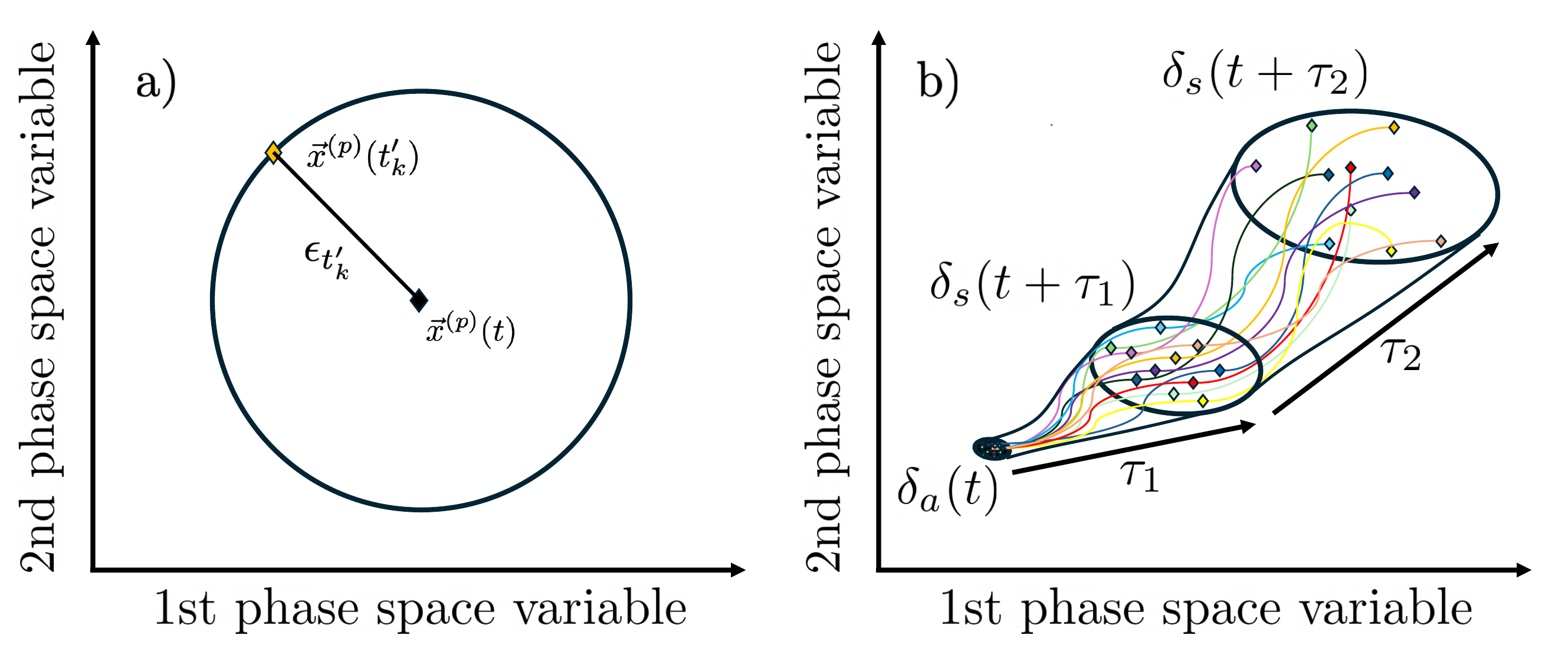}
 \centering
 \caption{a) Diagram of the formation of an ensemble of analogs $\mathbf{x_{a}}(t)$ in a $2$-dimensional reconstructed phase space from a given observation $\vec{x}^{(p=2)}(t)$. Considering the Euclidean distance, the cercle of center $\vec{x}^{(p=2)}(t)$ and radious $\epsilon_{t'_k}$ contains all the element of the analogs ensemble. The radious $\epsilon_{t'_k}$ is defined as the distance between the observation and its $k$th closest neighbor $\vec{x}^{(p=2)}(t'_k)$. b) Diagram of the time evolution of the analog ensemble in a $2$-dimensional reconstructed phase space. The smallest black ellipse $\delta_a(t)$ represents the volume occupied by analogs at time $t$. The two larger ellipses $\delta_s(t+\tau_1)$ and $\delta_s(t+\tau_2)$ represent respectively the volume occupied by successors at times $t+\tau_1$ and $t+\tau_2$. Coloured diamonds are the analog and succesor states, while coloured lines depict the trajectories followed by the analogs in time.}\label{fig:schema}
\end{figure}

Given a state $\vec{x}^{(p)}(t)$ in the $p$-dimensional phase space, we search its $k$ closest analogs, \textit{i.e.} its $k$ nearest neighbors in the phase space, to generate an ensemble $\mathbf{x_{a}}(t)=\{ \vec{x}^{(p)}(t'_i) \}_{1\le i\le k}$ of $k$ close states with associated successors $\mathbf{x_{s}}(t+\tau)=\{ x^{(q)}(t'_i+\tau) \}_{1\le i\le k}$, see figure~\ref{fig:schema}. This analog ensemble definition implies a threshold $\epsilon_{t'_k}$ defined as the distance between the observed state and its $k$th nearest neighbor, $\epsilon_{t'_k}=D(\vec{x}^{(p)}(t),\vec{x}^{(p)}(t'_k))$. Consequently, the threshold depends on the observed state and thus on the region of the phase space where it lies. In practice, we use past realizations of $X$ to look for analogs~\cite{lorenzAtmosphericPredictabilityRevealed1969a}, and so $t'_i<t$. If we have access to long enough realizations of $X$, we can consider that the $p$-dimensional phase space is densely sampled and so the $k$ analogs are extremely close.

There is no specific relation between the values of $p$ and $q$. $p$ is the dimension of the reconstructed phase space where one looks for analogs and it is suppossed to impact the analogs research. Each found analog corresponds to a state of the original 1-dimensional stochastic process $X$. Consequently, one can follow the time evolution of the analogs in a new reconstructed space of any dimension $q$ which can be equal to $p$ or not.

The analogs ensemble covers a volume $\delta_{a}(t)$ of the phase space. We statistically characterize the size of this volume, that can be understood as a separation between trajectories at time $t$, with the covariance matrix of the ensemble of analog sequences, more particularly:

\begin{equation} \label{eq:deltaa}
\delta_{a}(t) = |\Sigma_{\mathbf{x_{a}}(t)}|^{1/p}
\end{equation}

\noindent where $\Sigma_{\mathbf{x_{a}}(t)}$ is the covariance matrix and $| \, |$ is the determinant. In the case of a $1$-dimensional phase space ($p=1$), the size $\delta_{a}(t)$ is characterized by the variance $\mathbb{V}\left(\mathbf{x_{a}}(t)\right)$ of the analogs.

In the same way, the ensemble of successors at time $t+\tau$ covers a volume $\delta_{s}(t+\tau)$, that is statistically defined by:

\begin{equation} \label{eq:deltas}
\delta_{s}(t+\tau) = |\Sigma_{\mathbf{x_{s}}(t+\tau)}|^{1/q}
\end{equation}

Consequently, we can use the ensembles of analogs and successors to study how the the volume occupied by the ensemble evolves in time, \textit{i.e.} how the analogs disperse in the reconstructed phase space, see figure~\ref{fig:schema}. We can also study how the volume occupied by the successors depends on the volume occupied by the analogs: how the initial separation of trajectories at time $t$, that will mainly depend on the region of the phase space being studied, impacts the separation of trajectories at future times $t+\tau$. This, will be informative of the intrinsic dynamics of the studied stochastic process.

\section{Experimental turbulent velocity and scale-invariant stochastic processes}
\label{sec:data}

We study the dispersion of trajectories of analog-based ensembles in three different turbulent processes: a experimental turbulent velocity signal and two multifractal stochastic processes generated to mimic some of the statistics of turbulence.

\subsection{Experimental turbulent velocity}

We use an Eulerian longitudinal velocity measurement from an experimental grid turbulence setup in the Modane wind tunnel~\cite{Kahalerras1998}. This measurement was obtained using hot-wire anemometry and is $1$-dimensional. The Taylor scale Reynolds number of the flow is $R_\lambda = 2500$ and its mean velocity is $V=20.5$ m/s. The sampling frequency is $f_s=\frac{1}{dt}=25$ kHz.
From Taylor's frozen turbulence hypothesis~\cite{Frisch1995}, we can relate temporal and spatial variations. 
From previous studies, the integral and Kolmogorov time scales of the flow are respectively $T = 2350 dt$ and $\tau_K = 5 dt$~\cite{GraneroBelinchon2016}. For more details on the statistical properties of this turbulent velocity measurement see~\cite{Kahalerras1998, Froge2025, GraneroBelinchon2018}. 

\subsection{Scale-invariant stochastic processes}

We study a regularized fractional Brownian motion and a regularized Multifractal Random Walk~\cite{Robert2008,Chevillard2010, Pereira2016, GraneroBelinchon2024} parameterized to mimic Modane experimental turbulent velocity. Both $1$-dimensional processes can be modelled with the following stochastic integral~\cite{Robert2008, Pereira2016}:

\begin{equation}
X_{\mathcal{H},c_2,\tau_K}(z) = \int_{\mathbb{R}} \psi_T(z-z^\prime) P_{\mathcal{H},\tau_K} (z-z^\prime) M_{c_2,\tau_K}(z^\prime)W(z^\prime)d z^\prime
\label{def:field}
\end{equation}

\noindent where $z$ and $z^\prime$ denote $1$-dimensional position vectors, $W(z^\prime)$ is a Gaussian white noise, $0<\mathcal{H}<1$ the Hurst exponent and $c_2$ the intermittency coefficient. The first term, $\psi_T(z-z^\prime)$ is the large scale cut-off. The second term, $P_{\mathcal{H},\tau_K}(z) = \frac{1}{||z||_{\tau_K}^{1/2-\mathcal{H}}}$, is a kernel providing a power spectrum with a power law behavior of exponent $2\mathcal{H}+1$. The norm $||z||_{\tau_K} = \sqrt{||z||^2+\tau_K^2}$ in the denominator is a regularized $L^2$-norm with $\tau_K>0$ being the regularization scale and $||.||$ the $L^2$-norm.
The third term, $M_{c_2,\tau_K}(z^\prime)$ is a multiplicative chaos \cite{Robert2008} defined as:

\begin{equation}
M_{c_2,\tau_K}(z^\prime)=e^{-\sqrt{c_2} X_{\tau_K}(z^\prime)-c_2\mathbb{E}\{X_{\tau_K}^2(z^\prime)\}}
\end{equation}
\noindent where $X_{\tau_K}(z^\prime)$ is a log-correlated Gaussian noise with autocovariance function:

\begin{equation}\label{eq:autocov}
\mathbb{E} \{X_{\tau_K}(z)X_{\tau_K}(z^\prime) \}  \underset{||z-z^\prime||_{\tau_K}\rightarrow 0}\sim -\log(||z-z^\prime||_{\tau_K})
\end{equation}

When $c_2=0$, the process $X_{\mathcal{H},c_2=0,\tau_K}(z)$ is a r-fBm, which is a monofractal stationary stochastic process with Gaussian probability distribution. The statistical properties of the r-fBm are fully defined by its Hurst exponent $\mathcal{H}=1/3$ and the structure of covariance correctly mimics the one of turbulent velocity~\cite{Chevillard2012}. However, r-fBm does not display intermittency and consequently the probability distributions of its increments remain Gaussian across scales \textit{i.e.} small scales do not present extreme events. 

When $c_2 > 0$, the process $X_{\mathcal{H},c_2,\tau_K}(z)$ is a r-MRW, which is a multifractal stationary stochastic process with Gaussian probability distributions at large scales and heavy tailed probability distributions for scales in the inertial and dissipative domains. The statistics of r-MRW are prescribed by the log-normal multifractal model of turbulence~\cite{Chevillard2012} and so they are fully defined by its Hurst exponent and intermittency parameters that we fix respectively to $\mathcal{H}=1/3$ and $c_2=0.025$.  

For both processes, the Kolmogorov and integral scales are respectively fixed to $T = 2350 dt$ and $\tau_K = 5 dt$ as for Modane. This implies a Reynolds number for these processes of $R_\lambda \sim 2500$. The used large scale regularization function $\psi_T$ is Gaussian.

\section{Numerical implementation of the analog-based ensemble generation} 
\label{sec:num}

Assuming a stochastic process $X=\{x(t)\}$ sampled at intervals $dt$, we use a realization of lenght $T_{D}$ as historical database where one looks for analogs. In the following, we call it \emph{database}. Another independent realization of length $T_M$ is used to sample the phase space. It is called \emph{measure} in the following. For each state of the stochastic process in the measure, we will generate analog-based ensembles of states belonging to the database and we will study how these ensembles disperse in time. It is important to have a long enough database, in order to densely cover the full phase space. 

For any state at time $t$ of the measure realization, we estimate the volume $\delta_{a}(t)$ occupied by its analogs and the volume $\delta_{s}(t+\tau)$ occupied by their successors as the time $\tau$ evolves. Consequently, for a measure realization of length $T_M$ containing $N_M=\frac{T_M}{dt}$ samples, we have sets of $\delta_{a}(t)$ and $\delta_{s}(t+\tau)$ of the same size $N_M$.

For the three studied processes presented in section~\ref{sec:data}, we define a database and a measure realization. 
The full experimental turbulent velocity signal spans $1000$ seconds that corresponds to $10638$ integral scales. The first $419$ seconds, containing $5\times 2^{21}$ samples, define the database where looking for analogs. This database contains approximately $4460$ integral scales. The measure realization is defined for times $t \in [838,922]$ seconds, corresponding to $N_M=2^{21}$ samples. Leaving $419$ seconds between the database and the measure is useful to ensure that no correlations remain between them.
Equivalently, for both stochastic processes r-fBm and r-MRW, a single realization of size $5\times 2^{21}$ samples was generated to define the database where looking for analogs, and a realization of size $N_{M}=2^{21}$ samples was generated to define the measure. In this way, the experiments are performed equivalently for the Modane turbulent velocity and these two synthetic processes.

Once the database and the measure realizations are defined, the only three parameters to be fixed are: the dimension of the analogs $p$, the dimension of the successors $q$, and the number $k$ of trajectories in the analog ensemble.

The dimensions of analogs and successors are fixed respectively to $p=3$ and $q=1$ following the work of Frog\'e et al.~\cite{Froge2025}. On the one hand, the choice of $p=3$ to define the reconstructed phase space where one looks for analogs is supported by the three-point closure approximation for turbulence statistics from~\cite{Peinke2019} and by the full study of $p$ and $k$ parameters in forecast applications on Modane wind tunnel experimental signal from~\cite{Froge2025}. On the other hand, the choice of $q=1$ for the successors dimensionality is mainly supported by technical arguments: once the analogs are correctly identified in the $p$-dimensional phase space, we can just study how the corresponding $1$-dimensional projection evolves in time. The number of analogs $k$ has to be considerably larger than the dimension of the reconstructed phase space~\cite{Froge2025}.

All the results illustrated in this work are obtained with $k=200$, $p=3$ and $q=1$. Same qualitative results have been obtained with $(k=50, p=3, q=1)$, $(k=100, p=3, q=1)$ and $(k=50,p=3,q=3)$ with only slight differences due to statistical effects.

\section{Results}
\label{sec:results}

In this section, we study the evolution of the distance between trajectories, $\delta_s(t+\tau)$, in function of the time, $\tau$, and the initial separation, $\delta_a(t)$, for the three processes presented in section~\ref{sec:data}.

\begin{figure*}[ht]
 \includegraphics[width=1\linewidth]{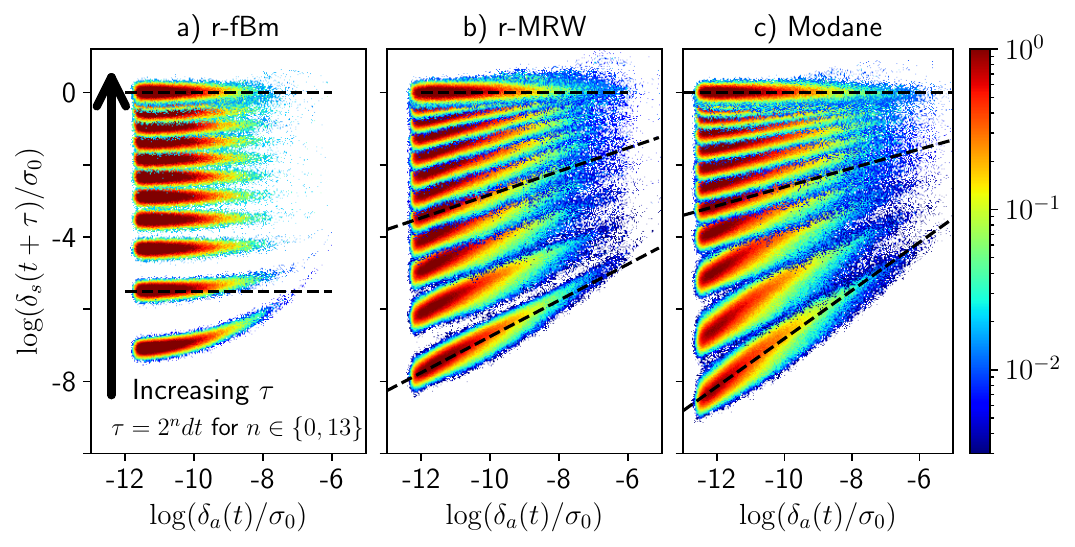}
 \centering
 \caption{Logarithm of the volume of the phase space occupied by successors $\delta_s(t+\tau)$ in function of the logarithm of the initial volume occupied by the analog states $\delta_a(t)$ for different values of $\tau$ and for a) r-fBm, b) r-MRW and c) experimental turbulent velocity measurement. A total of $14$ histograms are shown for each process corresponding respectively to $\tau=2^{n} dt$ with $n \in \left\lbrace 0, 13 \right\rbrace$. The variance of the studied process is $\sigma_0$.}\label{fig1}
\end{figure*}

Figure~\ref{fig1} shows the logarithm of the volume occupied by the ensemble of successors $\log \left( \delta_{s}(t+\tau) \right)_{t \in [0,T_m]}$ in function of the logarithm of the volume occupied by the ensemble of analogs $\log \left( \delta_{a}(t) \right)_{t \in [0,T_m]}$ for a) r-fBm, b) r-MRW and c) Modane experimental turbulence. Histograms are build on the set of all the states $t \in [0,T_m]$ of the measure realization, so each histogram contains $N_M=2^{21}$ volumes. Each histogram corresponds to a different time delay $\tau$ of the successors where $\tau=2^{n} dt$ with $n \in \left\lbrace 0 , 13 \right\rbrace$. We observe that $\delta_{s}(t+\tau)$ increases with $\tau$ regardless of the process being studied. Moreover, the manner of this increase is similar across all three processes. We also observe that in the case of r-fBm, $\delta_{s}(t+\tau)$ does not depend on $\delta_{a}(t)$ except very slightly at very small $\tau$. On the contrary, for r-MRW and the experimental turbulent velocity, $\delta_{s}(t+\tau)$ behaves as a power law of $\delta_{a}(t)$:

\begin{equation}\label{eq:alpha}
\delta_{s}(t+\tau) \sim \left( \delta_{a}(t) \right)^{\alpha(\tau)}
\end{equation}

\noindent with a power $\alpha(\tau)$ that depends on the time delay.

\begin{figure}[ht]
 \includegraphics[width=\linewidth]{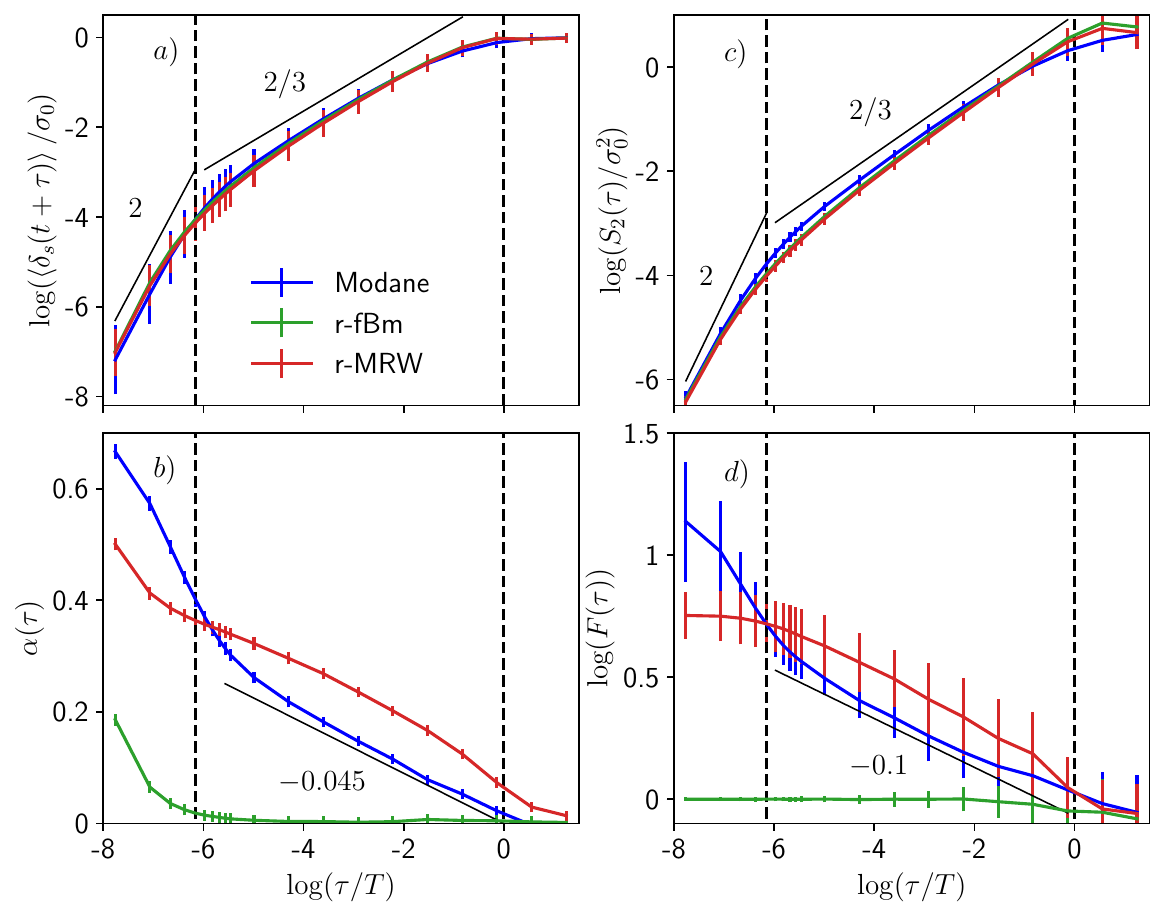}
 \centering
 \caption{Evolution of a) the logarithm of the average volume of the phase space occupied by successors at time $\tau$, $\log(\left\langle \delta_{s}(t+\tau) \right\rangle/\sigma_0)$, b) the exponent $\alpha(\tau)$ of the power law relationship between the volume occupied by successors at time $\tau$ and the initial volume occupied by analogs, c) the second order structure function and d) the flatness, all in function of $\log(\tau/T)$ for r-fBm in green, r-MRW in red and Modane experimental turbulent velocity in blue. Black dashed vertical lines indicate the integral $T$ and dissipative $\tau_K$ scales of the flow. The black straight lines in a) and c) have slopes of $2$ and $2/3$. The black straight lines in b) and d) have a slope of $-0.045$ and $-0.1=-4c_2$ respectively. The variance of the studied process is $\sigma_0$.}\label{fig2}
\end{figure}

Figure~\ref{fig2} a) shows the logarithm of the average of $\delta_{s}(t+\tau)$ in function of the time delay $\tau$, where the average is estimated over the set of states of the measure realization: $\left\langle \delta_{s}(t+\tau) \right\rangle = \frac{1}{N_M}\sum_{j=1}^{N_M}\delta_{s}(t_{j}+\tau)$. The three processes present exactly the same behavior: the average of the volume occupied by the successors $\left\langle \delta_{s}(t+\tau) \right\rangle$ increases with $\tau$ following power laws:

\begin{numcases}{\left\langle \delta_{s}(t+\tau) \right\rangle \sim }
    \tau^2     &for  $\tau<\tau_K$ \nonumber \\
    \tau^{2/3} &for $\tau_K<\tau<T$ \label{eq:avg} \\
    0          &for $\tau > T$ \nonumber
\end{numcases}

Moreover, this behavior perfectly matches the one of the second-order structure function of turbulent velocity~\cite{Frisch1995}, which is mimicked by r-fBm and r-MRW. The average dispersion of analog-based ensembles in function of time is governed by second order statistics, and more particularly by the distribution of energy across scales, see appendix~\ref{appendixA}. Similar results were obtained by Bitane et al.~\cite{Bitane2013} when studying velocity differences between pairs of Lagrangian tracers.

Figure~\ref{fig2} b) shows the evolution of $\alpha(\tau)$ in function of $\tau$ for the three processes. As observed before, for r-fBm $\alpha(\tau)=0$ independently of $\tau$, except for small $\tau$ in the dissipative domain, $\tau<\tau_K$. This means that $\delta_s(t+\tau)$ is independent of $\delta_a(t)$ for scales in the inertial and integral domains and slightly depends on it at dissipative scales~\cite{Bitane2013}. In the case of the r-MRW and the experimental turbulent velocity, $\alpha(\tau)$ is zero for time delays in the integral domain, $\tau>T$, and increases linearly when $\tau$ decreases through the inertial domain. In the dissipative domain the increase of $\alpha(\tau)$ with decreasing $\tau$ is steeper. These observations allow us to state that the dependence of $\delta_{s}(t+\tau)$ on $\delta_{a}(t)$ is governed by intermittency effects, \textit{i.e.} by extreme events at small scales in the inertial and dissipative ranges. The weak dependance of $\alpha(\tau)$ on $\tau$ observed for r-fBm at dissipative scales could correspond to chaotic effects of lesser importance than those driven by intermittency.

Figure~\ref{fig2} c) and d) show respectively the logarithm of the second order structure function $\log(S_2(\tau))$ and the logarithm of the flatness $\log(F(\tau))$ in function of the scale $\tau$ for the r-fBm, r-MRW and the Modane turbulent experimental signal. For any stochastic process $X=\{x(t)\}$, the second order structure function and the flatness are defined:

\begin{eqnarray}
S_2(\tau) &= \left\langle \left( x(t+\tau) - x(t) \right)^2 \right\rangle \\
F(\tau) &= \frac{\left\langle \left( x(t+\tau) - x(t) \right)^4 \right\rangle}{ \left\langle \left( x(t+\tau) - x(t) \right)^2 \right\rangle^2}
\end{eqnarray}

\noindent where $ \left\langle \, \right\rangle$ indicates the average that can be done across realizations or along the time dimension through ergodicity assumption. While the second order structure function is a measure of the energy distribution across scales, the flatness characterizes intermittency~\cite{Frisch1995}. Figure~\ref{fig2} illustrates that for all the three processes, $\left\langle \delta_{s}(t+\tau)/\sigma_0 \right\rangle$ characterizes the energy distribution across scales in the same way as $\log(S_2(\tau)/\sigma_0^2)$, and $\alpha(\tau)$ characterizes intermittency in the same way as $\log(F(\tau))$.

\section{Discussion}
\label{sec:conclusion}

In this article, we introduced a methodology for analyzing the dispersion of trajectories of stochastic processes in reconstructed phase spaces derived from observed data. The approach identifies ensembles of analog states, \textit{i.e.} states that are close in phase space, allowing us to examine how these states diverge over time. Specifically, we investigated the dispersion of analog states as a function of both the time delay and their initial separation. We studied an experimental Eulerian turbulent velocity signal, as well as two synthetic scale-invariant stochastic processes: a r-fBm and a r-MRW. Both synthetic processes are designed to replicate the covariance structure of the turbulent velocity signal, but only the multifractal random walk exhibits intermittency. 

Our results demonstrate that while the covariance structure determines the time dependence of analog dispersion, it is the presence of intermittency that drives the sensitivity of dispersion to the initial separation between analog states. More particularly, we illustrated that, for the three studied processes, the average of the variance of the ensemble of successors, \textit{i.e.} the average volume occupied by the successors states, behaves as a power law of $\tau$ following (\ref{eq:avg}). We also showed that, for the r-MRW and the experimental turbulent velocity, the dependence of $\delta_{s}(t+\tau)$ on the initial separation of the analogs follows (\ref{eq:alpha}) illustrating the impact of the small-scale extreme events on the dispersion of analog trajectories on the reconstructed phase space. On the contrary, for r-fBm, $\delta_{s}(t+\tau)$ is independent of the initial separation of the analogs except for small $\tau$. Comparable qualitative conclusions were found by Bitane et al.~\cite{Bitane2013} in their analysis of velocity differences between pairs of Lagrangian tracers. 

Analog methods~\cite{lorenzAtmosphericPredictabilityRevealed1969a} allows us to define ensembles of similar $p$-dimensional states, which serve as a basis for analyzing the dynamical properties of the studied process, particularly the evolution of the distance between trajectories as a function of their initial separation and time delay. In Appendix~\ref{appendixB}, we present the results obtained using randomly sampled states instead of analogs, which clearly demonstrates the importance of identifying true analog states. 

In this work, we defined the phase space through Taken's embedding and the volume occupied by the ensembles with their covariance matrix, see (\ref{eq:deltaa}) and (\ref{eq:deltas}). However, the analog-based approach is generic and can be used with different phase space definitions and different ensemble volume estimators. Indeed, replacing the volume definition in (\ref{eq:deltaa}) and (\ref{eq:deltas}) by the average of the Euclidean distance between pairs of ensemble states leads to the same qualitative results.

The methodology presented in this article exhibits several limitations. First of all, it grounds on the Poincar\'e recurrence theorem and so considers that one can find analog states in the database. However, the  kac's lemma implies that to find analogs the size of the needed database $N_M$ must scale as $N_M \sim \left( \frac{L}{\epsilon} \right)^{d_A}$ with $d_A$ the dimension of the attractor of the system, $\epsilon$ the threshold in the analog search and $L$ a characteristic length-scale of the attractor~\cite{Cecconi2012,Chibbaro2014}. Consequently, high-dimensional systems, such as turbulence, are specially difficult to study with the proposed methodoloy. In this work, for all the studied processes, we use databases with $N_M=5\times 2^{21}$ samples, each one containing approximately $4460$ integral scales of the flow. Consequently, we have access to very huge databases that are not always available in applications such as climate or weather~\cite{Noyelle2025}. In addition, we use Takens embedding with dimension $p=3$ to reconstruct the phase space where one looks for analogs. This implies a contraction of the phase space of turbulence to a 3-dimensional one, that could lead to an increase of the density of states in the reconstructed phase space. Finally, contrary to other studies such as~\cite{Cecconi2012}, we don't fix the threshold distance $\epsilon$ characterizing the analogs but we use a state-dependent definition $\epsilon_{t'_k}$ which allows to ensure the finding of analogs. Evidently, the quality of the found analogs will depend on the region of the phase space, its dynamics and how densely it has been sampled in the database.

Future research will focus on the investigation of different phase space reconstructions. On the one hand, by studying the impact of the dimension and time delay in Takens embedding. In this work, we fixed the dimension to $p=3$ and the time delay to the sampling distance $dt$, but other choices such as $\tau_K$ for the delay could be of interest. On the other hand, by exploring new statistical approaches for phase space reconstruction~\cite{Healy2024,Maaten2008}. Other perspective is the study of shell models of turbulence, which are dynamical systems displaying energy cascade and intermittency~\cite{Mailybaev2013,Aumaitre2024}. The study of these toy models can help to better understand analog dispersion in intermittent systems. Moreover, the feasability of the methodology should be validated on turbulent databases of higher dimension. Other research perspectives are: the study of the influence of the Reynolds number on $\delta_s({t+\tau})$ through the analysis of experimental turbulent velocity data~\cite{Chanal2000} and the characterization of the impact of the Hurst exponent and the intermittency parameters on the dispersion of analog ensembles of scale-invariant synthetic stochastic processes.

\acknowledgments

The author has no interests to declare. The author wishes to thank S. G Roux for providing the code used for the generation of scale-invariant stochastic processes that is freely available at : https://gitlab.com/sroux67/multivariate-multifractal-field-synthesis. This work was supported by the French National Research Agency under Grant ANR-21-CE46-0011-01, within the program ``Appel \`a projets g\'en\'erique 2021''.

\appendix

\section{Analytical relation between variance of successors and variance of increments}\label{appendixA}

Given a centered stationary stochastic process $X$, we define its increment of size $\tau$ as:

\begin{equation}
\Delta_{\tau}X(t) = X(t)-X(t-\tau)
\end{equation}

For a given state $x(t)$ of $X$, we find an ensemble of $k$ analogs and their respective successors and we write the variance of the ensemble of successors as:

\begin{align}
\delta_s(t+\tau) &= \frac{1}{k} \sum_{i=1}^{k} x^2(t_i+\tau) \\
&= \frac{1}{k} \sum_{i=1}^{k} (x(t_i)+\Delta_{\tau}X(t_i+\tau))^{2}  \\
&= \frac{1}{k} \sum_{i=1}^{k} x^2(t_i) + \frac{1}{k} \sum_{i=1}^{k} \left( \Delta_{\tau}X(t_i+\tau)\right)^{2} + \\
& \qquad + \frac{2}{k} \sum_{i=1}^{k} x(t_i) \Delta_{\tau}X(t_i+\tau) \nonumber \\
&= \mathbb{V}(X(t)) + \mathbb{V}\left(\Delta_{\tau}X(t)\right) + \\
& \qquad +2 R(X(t), \Delta_{\tau}X(t+\tau)) \nonumber
\end{align}


\noindent where $R(X(t), \Delta_{\tau}X(t+\tau))$ is the cross-correlation between the process and its next increment. In the case of turbulent velocity, $R(X(t), \Delta_{\tau}X(t+\tau))\approx 1$ for $\tau<\tau_{K}$ and $\approx 0$ for $\tau\geq T$.  

Consequently, the behavior of $\delta_s(t+\tau)$ in function of $\tau$ should be different that the one of $\mathbb{V}\left(\Delta_{\tau}X(t)\right)$. However, if we make the assumption that the $k$ analog states are identical to the state $x(t)$, we can consider that $x(t_i)=x(t)$ independent of $i$. This leads to:

\begin{align}
\delta_s(t+\tau) &= \frac{1}{k} \sum_{i=1}^{k} x^2(t_i) + \frac{1}{k} \sum_{i=1}^{k} \left(\Delta_{\tau}X(t_i+\tau)\right)^{2} + \\
& \qquad + \frac{2}{k} \sum_{i=1}^{k} x(t_i) \Delta_{\tau}X(t_i+\tau) \nonumber\\
&= x^2(t) + \frac{1}{k} \sum_{i=1}^{k} \left(\Delta_{\tau}X(t_i+\tau)\right)^{2} + \\
& \qquad + 2x(t) \frac{1}{k} \sum_{i=1}^{k} \Delta_{\tau}X(t_i+\tau) \nonumber \\
&= x^2(t) + \mathbb{V}\left(\Delta_{\tau}X(t)\right) + 2x(t) \mathbb{E}\left(\Delta_{\tau}X(t)\right)
\end{align}


Since the increments are centered and $x^2(t)$ is independent of $\tau$, we can approximate $\delta_s(t+\tau) \sim  \mathbb{V}\left(\Delta_{\tau}X(t)\right)$.

\section{Results with random sampling instead of analogs}\label{appendixB}

\begin{figure*}[ht]
 \includegraphics[width=1\linewidth]{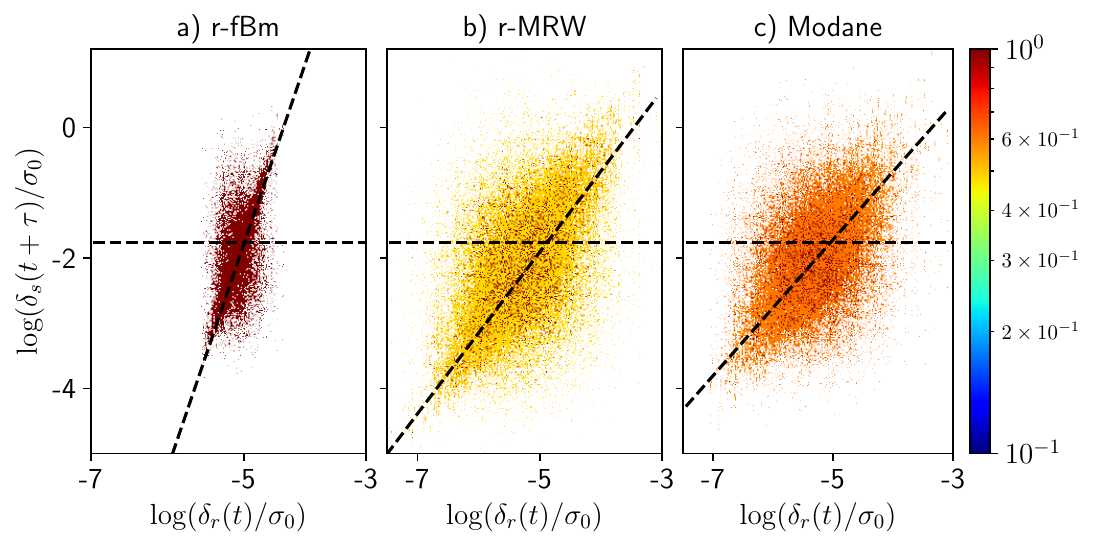}
 \centering
 \caption{Logarithm of the volume of the phase space occupied by successors $\delta_s(t+\tau)$ in function of the logarithm of the initial volume occupied by the randomly sampled states $\delta_r(t)$ for different values of $\tau$ and for a) r-fBm, b) r-MRW and c) experimental turbulent velocity measurement. A total of $14$ histograms are shown for each process corresponding respectively to $\tau=2^{n} dt$ with $n \in \left\lbrace 0, 13 \right\rbrace$. The variance of the studied process is $\sigma_0$.}\label{fig3}
\end{figure*}

\begin{figure}[ht]
 \includegraphics[width=0.5\linewidth]{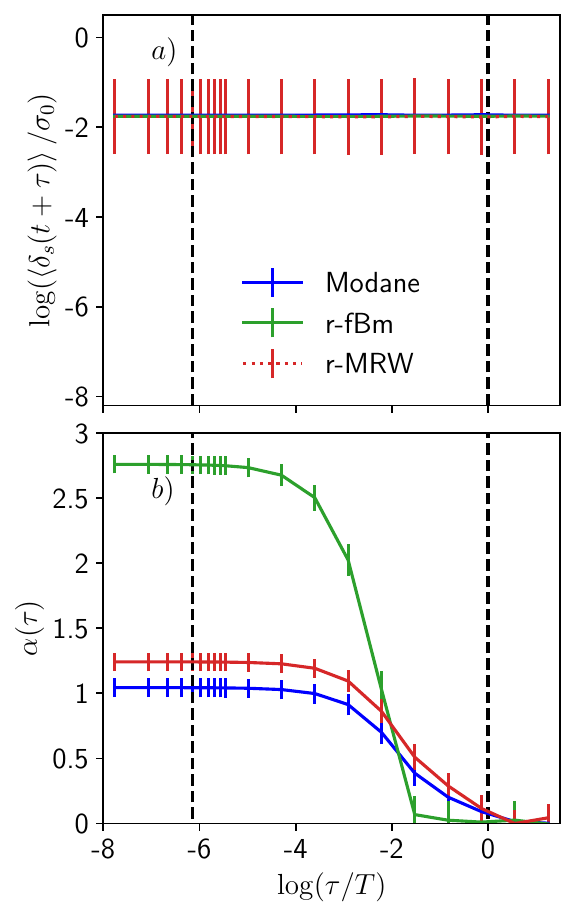}
 \centering
 \caption{Evolution of a) the logarithm of the average volume of the phase space occupied by successors at time $\tau$, $\log(\left\langle \delta_{s}(t+\tau) \right\rangle/\sigma_0)$, and b) the exponent $\alpha(\tau)$ of the power law relationship between the volume occupied by successors at time $\tau$ and the initial volume occupied by randomly sampled states, both in function of $\log(\tau/T)$ for r-fBm in green, r-MRW in red and Modane experimental turbulent velocity in blue. In both a) and b), black dashed vertical lines indicate the integral $T$ and dissipative $\tau_K$ scales of the flow. The variance of the studied process is $\sigma_0$.}\label{fig4}
\end{figure}

In this section, we illustrate the benefit of using analogs to define the initial position of the studied trajectories in the phase space. If instead of defining the initial volume occupied by analogs $\delta_a(t)$, we use an initial volume occupied by randomly sampled states of the phase space $\delta_r(t)$ the temporal dependence of the dispersion of the trajectories is lost. This can be seen in figure~\ref{fig3} and~\ref{fig4} a).

In figure~\ref{fig4} b), we show the behavior of the exponent $\alpha(\tau)$ estimated for the relatioship $\delta_s(t+\tau)\sim \delta_r(t)^{\alpha(\tau)}$ in function of $\log(\tau/T)$. Now, all the three processes present $\alpha(\tau) \approx 0$ for large $\tau$ and a plateau at a process-dependent positive value for small $\tau$ in the low inertial and dissipative domains. The r-fBm has the highest plateau at $2.75$, while r-MRW and Modane present the plateau at $1.25$ and $1.05$ respectively. From figure~\ref{fig3}, we consider that these measured exponents have to be interpreted prudently since the clouds of points seems quite dispersed.

\clearpage

\bibliographystyle{elsarticle-num} 
\bibliography{biblio}

\end{document}